\documentclass[aps,pra,twocolumn,groupedaddress]{revtex4-1}
\usepackage{amsmath,amssymb,graphicx,bm}
\begin{document}
\title{Su-Schrieffer-Heeger-type Floquet network}
\author{Tetsuyuki Ochiai}
\affiliation{Research Center for Functional Materials, National Institute for Materials Science (NIMS), Tsukuba 305-0044, Japan}
\date{\today}

\begin{abstract}
A network model that can describe light propagation in one-dimensional ring-resonator arrays with a dimer structure is studied as a Su-Schrieffer-Heeger-type Floquet network. The model can be regarded as a Floquet system without periodic driving and exhibits quasienergy band structures of the ring propagation phase. Resulting band gaps support deterministic edge states depending on hopping S-matrices between adjacent rings. The number of edge states is one if the Zak phase is $\pi$. If the Zak phase is 0, the number is either zero or two. The criterion of the latter  number is given analytically in terms of the reflection matrix of the semi-infinite system.   These properties are directly verified by changing S-matrix parameters and boundary condition continuously.   
 \end{abstract}

\maketitle

\section{introduction}
Floquet engineering is a keyword in current condensed matter physics \cite{bukov2015universal}.  By irradiating electronic systems with monochromatic light, the systems can show exotic physical properties that are not accessible without the irradiation \cite{PhysRevB.79.081406,PhysRevLett.105.017401}. The systems are thus engineered by driving external fields. Theoretically,  such a system is described by a time-dependent and time-periodic hamiltonian, and is called a Floquet system.  Its eigenvalue equation is the diagonalization of the unitary time-translation operator for one period.

Interestingly, similar eigenvalue equations of unitary matrices are obtained in certain network models without disorder \cite{PhysRevB.89.075113,PhysRevB.95.205413} and in discrete-time quantum walks \cite{PhysRevB.86.195414}. Network models were first introduced to study the Anderson localization in quantum Hall systems \cite{0022-3719-21-14-008}, and were later applied to light propagation in ring-resonator lattices in two or three spatial dimensions \cite{hafezi2011robust,PhysRevLett.110.203904,PhysRevB.93.144114}. These investigation unveils various interesting topological properties of the network models including single Dirac cone at $\Gamma$ \cite{PhysRevB.54.8708,PhysRevLett.117.013902}, anomalous Floquet insulator phases \cite{PhysRevB.89.075113}, Floquet-Weyl phases \cite{PhysRevB.93.144114,ochiai2016floquet}, and synthetic gauge fields \cite{hafezi2011robust,0953-8984-29-4-045501}.
There, driving external fields are completely absent.  The network models (together with discrete-time quantum walks) share the same mathematical structures inherent in Floquet systems, so that resulting physical phenomena are quite similar to Floquet ones, and thus can be exotic and topological.  These properties enable us to study network models in more detail as alternative Floquet systems \footnote{Another alternative Floquet systems without time-periodic driving is the helix array of optical waveguides \cite{rechtsman2013photonic}. }.

Here, we propose a simple one-dimensional (1D) network model that imitates the Su-Schrieffer-Heeger (SSH) model \cite{PhysRevLett.42.1698}. The former model is a Floquet network in the sense given above. It can simulate the light propagation in 1D ring-resonator arrays. The latter model is a well-known tight-binding model that exhibits a topological phase transition regarding the Zak phase \cite{PhysRevLett.62.2747} and deterministic zero-energy edge states. One of the advantages in the former model is that we can change continuously the boundary condition, which is often crucial for edge states. We can thus directly check the robustness of possible edge states by changing the boundary condition.

On the contrary to a common understanding of the SSH model, in this paper on the 1D network model, we find edge states even if the Zak phase is zero. The common understanding of no edge state for the zero Zak-phase systems is limited in a certain region of possible boundary conditions. We derive these results analytically by considering a relation between the Zak phase and a S-matrix of the semi-infinite system with edge.  We also directly confirm the above prediction numerically.   

We should note that there is another route to realize 1D Floquet systems from the SSH model \cite{PhysRevA.92.023624}.  That is, just replacing the hopping parameters by those with Piers phases of driving laser light.  In contrast to this route, our model is easier to solve nonperturbatively and exhibits various different properties.

An optical realization of the present model seems to have less difficulty.  Using Silicon nanowire, it is possible to realize 1D ring-resonator arrays with high quality \cite{xia2007ultracompact}. Target frequencies can vary from visible to infrared. With such optical systems, novel phase modulations via topological band gaps, nonlinear optical effects via strongly localized edge states with tunable resonant frequencies, and topological lasing such as given in \cite{ota2018topological} 
will occur  using the properties discussed in this paper. Therefore, even in simpler 1D optical structures, various applications using topological effects are available.

This paper is organized as follows. In Sec. 2, we define a SSH-type network model. In Sec. 3, bulk eigenmodes are considered and the phase diagram regarding the Zak phase is derived. In Sec. 4, finite systems with edges are studied. Their eigenvalue spectra and analytic properties of a relevant S-matrix are investigated. The conditions of finding edge states are also given.  
 Finally, in Sec. 5, we summarize the results.

\section{model}

\begin{figure*}
\includegraphics[width=0.8\textwidth]{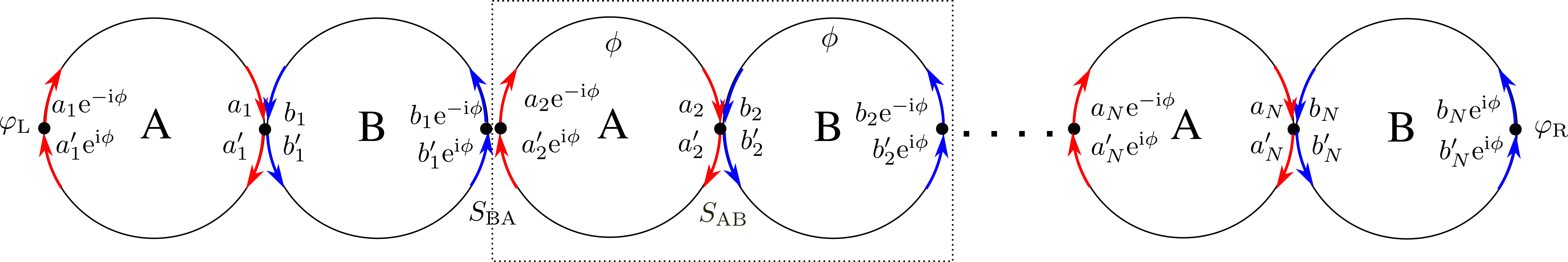}
\caption{\label{Fig_geo1d} Schematic illustration of the 1D SSH-type network model. It consists of identical rings, in which light flows alternately clockwise and counter-clockwise in A and B rings, respectively. Light amplitudes $a_n$, $a_n$, $b_n$, and $b_n'$ are introduced in the rings, and acquire phase $\phi$ per propagating one half of the ring. Phase delays $\varphi_{\rm L}$ and $\varphi_{\rm R}$ are introduced at the left and right boundary rings, respectively. The dashed region represents a unit cell. }
\end{figure*}

Let us consider a network model for the light propagation in the linear chain of ring resonator. A schematic illustration of the system under study is shown in Fig. \ref{Fig_geo1d}. 
We consider identical rings and control the distances between adjacent rings such that a dimer structure is formed. As a result, there are two rings (A and B rings) per unit cell. 
In the A rings, light flows clockwise, whereas in the B rings, light flows counter-clockwise. In  both the rings, the light propagating phase per half circle is common and denoted as $\phi$.

At the AB interface, we introduce the hopping S-matrix as 
\begin{align}
&\left(\begin{array}{c}
a_n'\\
b_n'
\end{array}\right)=S_{\rm AB}\left(\begin{array}{c}
a_n\\
b_n
\end{array}\right),\label{Eq_S_AB}\\
&S_{\rm AB}=\left(\begin{array}{cc}
r_{\rm AB} & t'_{\rm AB}\\
t_{\rm AB} & r'_{\rm AB}
\end{array}\right),  
\end{align}
where $S_{\rm AB}$ is unitary provided that there is no dissipation. 
Similarly, at the BA interface, the S-matrix is written as 
\begin{align}
&\left(\begin{array}{c}
a_{n+1}{\rm e}^{-{\rm i}\phi}\\
b_n   {\rm e}^{-{\rm i}\phi}
\end{array}\right)=S_{\rm BA}\left(\begin{array}{c}
a'_{n+1}{\rm e}^{{\rm i}\phi}\\
b'_n    {\rm e}^{{\rm i}\phi}
\end{array}\right),\label{Eq_S_BA}\\
&S_{\rm BA}=\left(\begin{array}{cc}
r_{\rm BA} & t'_{\rm BA}\\
t_{\rm BA} & r'_{\rm BA}
\end{array}\right). 
\end{align}
Again, $S_{\rm BA}$ is assumed to be unitary.

Because of the mirror symmetry between the A and B rings, the S-matrices satisfy 
\begin{align}
& S_{\rho}=\sigma_1 S_{\rho}\sigma_1 \quad (\rho={\rm AB},{\rm BA}), \\
& \sigma_1=\left(\begin{array}{cc}
0 & 1\\
1 & 0
\end{array}\right). 
\end{align} 
Therefore,  they are parametrized as 
\begin{align}
S_{\rho}={\rm e}^{{\rm i}\kappa_\rho}\left(\begin{array}{cc}
\cos\theta_\rho & {\rm i}\sin\theta_\rho \\
{\rm i}\sin\theta_\rho & \cos\theta_\rho 
\end{array}\right). 
\end{align}
The parameter $\theta_\rho$ represents the coupling strength between the A and B rings, and is controlled by the distance between the rings.  
The parameter $\kappa_\rho$ is the overall phase, and can be absorbed in the redefinition of the ring propagation phase $\phi$. In what follows,  we put $\kappa_\rho=0$.

We should note that if the dimer-like structure is absent, the system is merely a conventional model for coupled ring-resonator optical waveguides \cite{poon2004matrix}.  Nevertheless, the present system with the dimer-like structure exhibits rich topological structures that are not known in the context of photonics.

\section{bulk systems}

In the bulk system, the Bloch theorem can be applied. After introducing Bloch momentum $k_x$ as $a_{n+1}=a_n\exp({\rm i}k_x)$ etc, Eqs. (\ref{Eq_S_AB}) and (\ref{Eq_S_BA}) cast into the following equation: 
\begin{align}
&U_{\rm bulk}(k_x)\psi_{k_x}={\rm e}^{-2{\rm i}\phi}\psi_{k_x},\quad \psi_{k_x}=\left(\begin{array}{c}
a_n\\
b_n
\end{array}\right),\\
&U_{\rm bulk}(k_x)=\tilde{S}_{\rm BA}S_{\rm AB},\\
&\tilde{S}_{\rm BA}=\left(\begin{array}{cc}
{\rm e}^{-{\rm i}k_x} & 0 \\
0 & 1
\end{array}\right)S_{\rm BA} \left(\begin{array}{cc}
{\rm e}^{{\rm i}k_x} & 0 \\
0 & 1
\end{array}\right)
\end{align}  
This is an eigenvalue equation for unitary matrix $U_{\rm bulk}$ and the propagation phase $\phi$ acts as the compactified eigenphase. Therefore, the equation is analogous to that in the Floquet system with a time-periodic hamiltonian, where the diagonalization is for a time-translation operator of one period.

\begin{figure}
\includegraphics[width=0.45\textwidth]{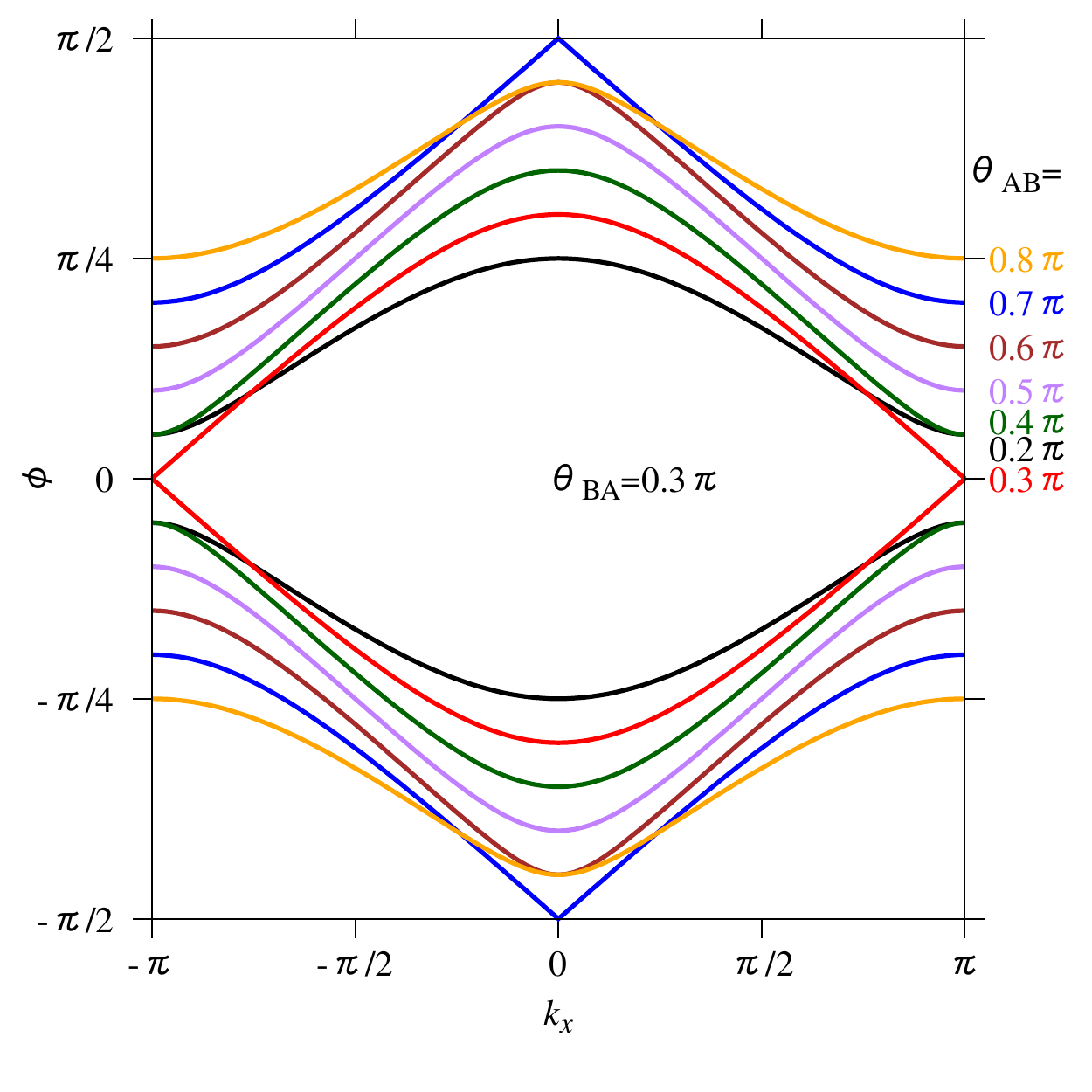}
\caption{\label{Fig_bulk1d} Quasienergy band structure of the 1D SSH-type Floquet network. We assume $\theta_{\rm BA}=0.3\pi$.  Parameter $\theta_{\rm AB}$ are changed from $0.2\pi$ to $0.8\pi$.  } 
\end{figure}

Figure \ref{Fig_bulk1d} shows typical quasi-energy band structures of the 1D SSH-type network model.
We have the two bands that are symmetric under $k_x$ and $\phi$ inversions. 
The band gaps are found around  $\phi=0$  and  $\phi=\pm \pi/2$.  
The former gap  closes at $\theta_{\rm AB}=\theta_{\rm BA}$, whereas the latter gap closes at $\theta_{\rm AB}+ \theta_{\rm BA}=\pm \pi$. 
At generic points of $(\theta_{\rm AB},\theta_{\rm BA})$, the band width $w$ is given by 
\begin{align}
w={\rm min}(|\theta_{\rm AB}|,|\theta_{\rm BA}|,\pi-|\theta_{\rm AB}|,\pi-|\theta_{\rm BA}|),
\end{align}    
provided $|\theta_{\rm AB}|,|\theta_{\rm BA}|\le \pi$.

The symmetry under  $k_x$ inversion is due to the mirror symmetry.  We can easily show   
\begin{align}
\sigma_1U_{\rm bulk}(k_x)\sigma_1 =U_{\rm bulk}(-k_x). 
\end{align}
Therefore, we have 
\begin{align}
&\psi_{-k_x}={\rm e}^{{\rm i}\lambda_{k_x}}\sigma_1\psi_{k_x}, \label{Eq_psi_inv}\\
&\phi(-k_x)=\phi(k_x).
\end{align}
We should note that the time-reversal symmetry is broken. This broken time-reversal symmetry is due to the decoupling between the clockwise and counter-clockwise modes in each ring, which is implicitly assumed in the formulation. 
If the contact region between adjacent rings is large enough compared to the relevant  wavelength of light, this assumption is fairly justified.

Besides, under the inversion of $\theta_{\rm AB}$ and $\theta_{\rm BA}$, we have a momentum shift as  
\begin{align}
\phi_{-\theta_{\rm AB},\theta_{\rm BA}}(k_x)=\phi_{\theta_{\rm AB},-\theta_{\rm BA}}(k_x)=\phi_{\theta_{\rm AB},\theta_{\rm BA}}(k_x+\pi). \label{Eq_1d_shift} 
\end{align} 
Also, under the swap of $\theta_{\rm AB}$ and $\theta_{\rm BA}$, we have a $k_x$ inversion:  
\begin{align}
&\phi_{\theta_{\rm BA},\theta_{\rm AB}}(k_x)
=\phi_{\theta_{\rm AB},\theta_{\rm BA}}(-k_x).  \label{Eq_1d_swap} 
\end{align} 
The inversion of both  $\theta_{\rm AB}$ and $\theta_{\rm BA}$ leads to the $\phi$-flip:  
\begin{align}
\phi_{-\theta_{\rm AB},-\theta_{\rm BA}}(k_x)=-\phi_{\theta_{\rm AB},\theta_{\rm BA}}(k_x). \label{Eq_1d_flip}
\end{align}
Combining Eqs. (\ref{Eq_1d_shift}) and  (\ref{Eq_1d_flip}), we have the chiral symmetry. Namely, the band structure is symmetric under the $\phi$ inversion as shown in Fig. \ref{Fig_bulk1d}.

\begin{figure}
\includegraphics[width=0.45\textwidth]{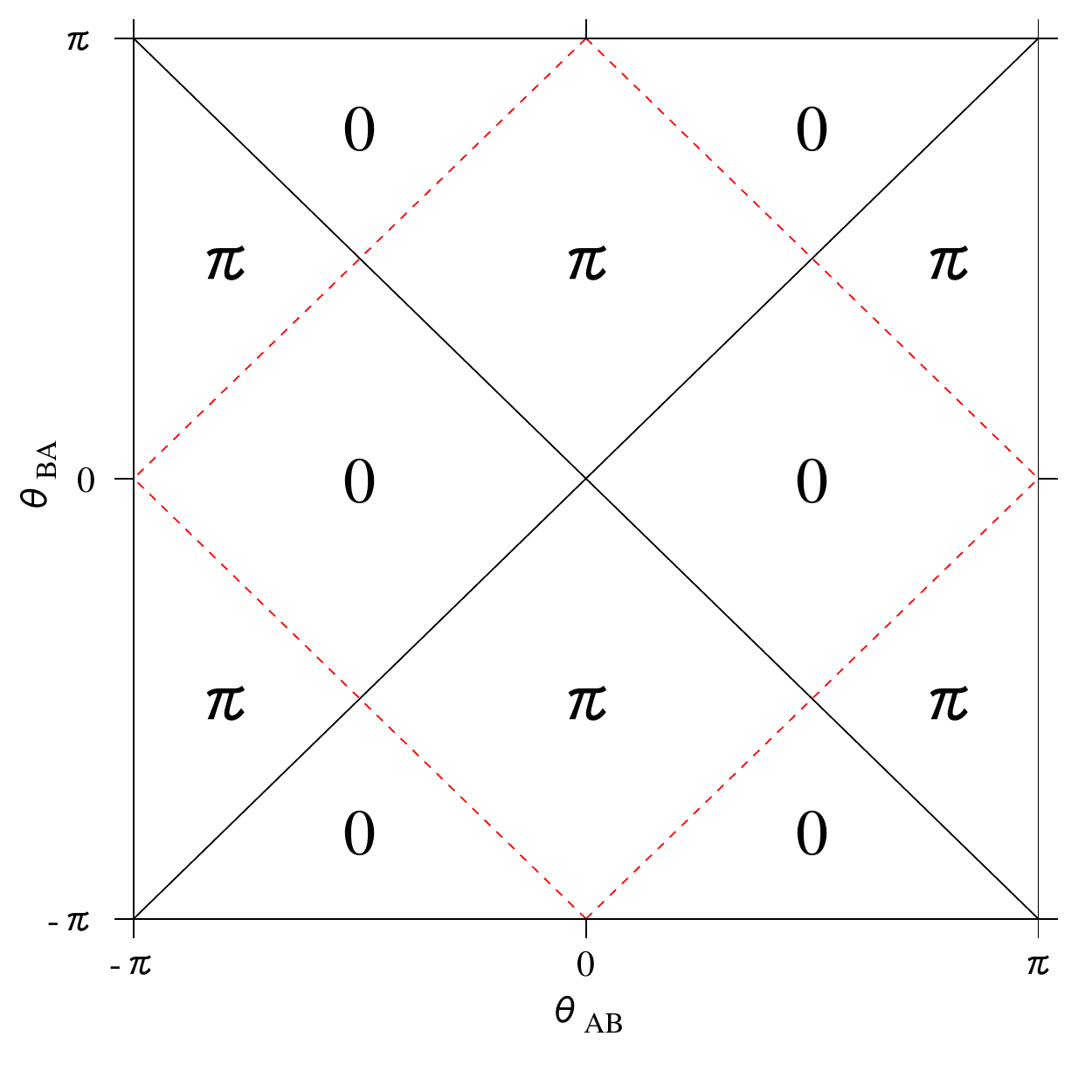}
\caption{\label{Fig_bulk1d_Zak} Phase diagram regarding the Zak phase, in the parameter space of $(\theta_{\rm AB},\theta_{\rm BA})$. On the solid line, the band gap around $\phi=0$ closes. On the dashed line, the band gap around $\phi=\pm\pi/2$ closes. These lines form the phase boundaries. }
\end{figure}

As in the SSH model, topological properties of the Floquet network are characterized by the Zak phase 
\begin{align}
\theta_{\rm Z} = \int_{-\pi}^\pi {\rm d}k_x \psi_{k_x}^\dagger {\rm i}\frac{\partial}{\partial k_x} \psi_{k_x} \quad {\rm mod}(2\pi), 
\end{align}
which is equal to either 0 or $\pi$. 
Using the mirror symmetry [Eq. (\ref{Eq_psi_inv})], we can easily show that 
\begin{align}
\theta_{\rm Z} = \lambda_\pi-\lambda_0. \label{Eq_Zak}
\end{align}
Both $\lambda_\pi$ and $\lambda_0$ take either 0 or $\pi$.  
The phase diagram regarding the Zak phase is shown in Fig. \ref{Fig_bulk1d_Zak}.   
Since both $\theta_{\rm AB}$ and $\theta_{\rm BA}$ are angle variables, the resulting phase diagram is a bit complicated compared with the SSH model.

\section{finite systems}

Next, let us consider a finite system with $2N$ rings. To study possible eigenmodes in the finite system, we need to fix the boundary condition. 
The boundary condition at the edge rings is taken as 
\begin{align}
&a_1={\rm e}^{2{\rm i}\phi}{\rm e}^{{\rm i}\varphi_{\rm L}}a'_1, \label{Eq_left_BC}\\
&b_N={\rm e}^{2{\rm i}\phi}{\rm e}^{{\rm i}\varphi_{\rm R}}b'_N. \label{Eq_right_BC}
\end{align}      
Here, we introduce the phase delays $\varphi_{\rm L}$ and $\varphi_{\rm R}$ there,  resulting in continuous control of the boundary condition. Such a controllability is one of the advantages in the network model.  In the tight-binding model, possible boundary conditions are limited and generally discrete.   In our case, however, we can choose freely the boundary condition, providing us a direct check of robustness in possible edge states.

The eigenvalue equation for the finite system is again the diagonalization of the unitary matrix:
\begin{align}
&U_{\rm finite}\left(\begin{array}{c}
{\bm a}\\
{\bm b}
\end{array}\right)={\rm e}^{-2{\rm i}\phi}\left(\begin{array}{c}
{\bm a}\\
{\bm b}
\end{array}\right),\\
&{\bm a}=(a_1,a_2,...,a_N)^t, \quad {\bm b}=(b_1,b_2,...,b_N)^t,\\ 
&U_{\rm finite}=\left(\begin{array}{cc}
U_{aa} &  U_{ab}\\
U_{ba} &  U_{bb}
\end{array}\right),\\
&[U_{aa}]_{nm}=\left\{\begin{array}{ll}
{\rm e}^{{\rm i}\varphi_{\rm L}}r_{\rm AB} & n=m=1\\
r_{\rm BA}r_{\rm AB} & n=m\ge 2\\
t'_{\rm BA}t_{\rm AB} & n=m+1\\
0 & {\rm otherwise}
\end{array}\right.,\\
&[U_{ab}]_{nm}=\left\{\begin{array}{ll}
{\rm e}^{{\rm i}\varphi_{\rm L}}t'_{\rm AB} & n=m=1\\
r_{\rm BA}t'_{\rm AB} & n=m\ge 2\\
t'_{\rm BA}r'_{\rm AB} & n=m+1\\
0 & {\rm otherwise}
\end{array}\right.,\\
&[U_{ba}]_{nm}=\left\{\begin{array}{ll}
{\rm e}^{{\rm i}\varphi_{\rm R}}r_{\rm AB} & n=m=N\\
r'_{\rm BA}t_{\rm AB} & n=m\le N-1\\
t_{\rm BA}r_{\rm AB} & n=m-1\\
0 & {\rm otherwise}
\end{array}\right.,\\
&[U_{bb}]_{nm}=\left\{\begin{array}{ll}
{\rm e}^{{\rm i}\varphi_{\rm R}}r'_{\rm AB} & n=m=N\\
r'_{\rm BA}r'_{\rm AB} & n=m\le N-1\\
t_{\rm BA}t'_{\rm AB} & n=m-1\\
0 & {\rm otherwise}
\end{array}\right..
\end{align}

\begin{figure}
\centerline{
\includegraphics[width=0.45\textwidth]{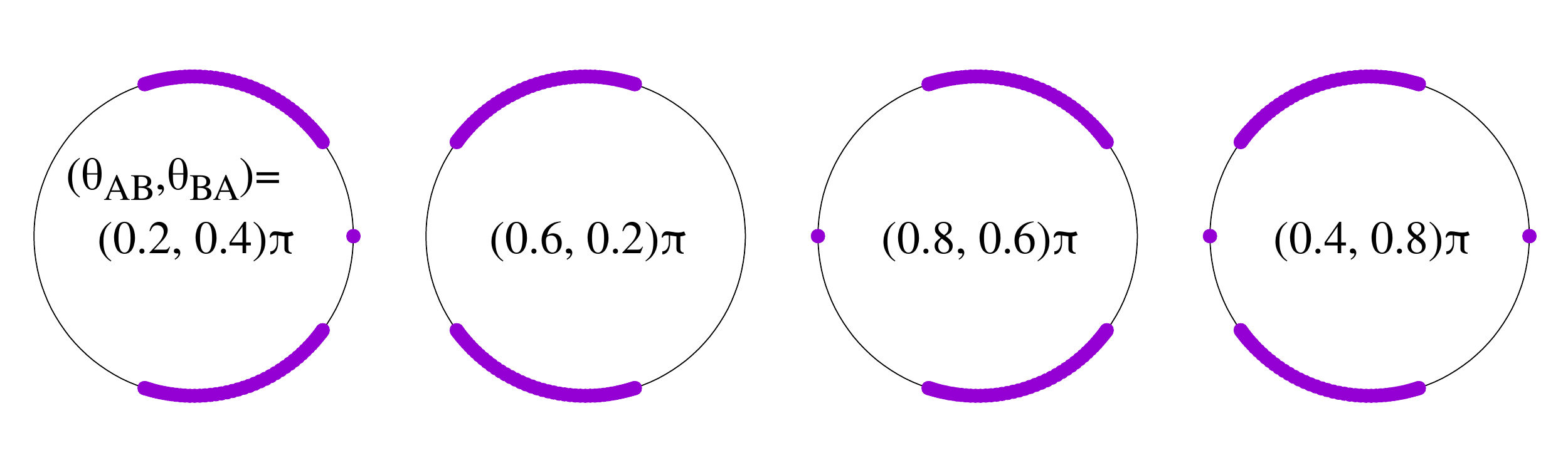}
}
\caption{\label{Fig_finite1d} Quasienergy eigenvalues in the finite systems of  $N=50$. Eigenvalues are plotted in the unit circle of $\exp(-2{\rm i}\phi)$ in the complex plane.  Four representative points of $(\theta_{\rm AB},\theta_{\rm BA})$ located in different regions in the phase diagram of Fig. \ref{Fig_bulk1d_Zak} are taken. }
\end{figure}

Typical eigenvalue spectra for the finite systems are shown in Figs. \ref{Fig_finite1d}.
 For simplicity, we assume zero phase delay $\varphi_{\rm L}=\varphi_{\rm R}=0$ at the edges. 
Depending on $(\theta_{\rm AB},\theta_{\rm BA})$, edge states emerge in a different way. For instance, at $(\theta_{\rm AB},\theta_{\rm BA})=(0.2\pi,0.4\pi)$, we have two edge state (bonding and antibonding orbitals of the left and right edges) at $\phi=0$ ,  whereas at $(0.6\pi,0.2\pi)$, no edge states are found.  Since the former and latter systems have $\theta_{\rm Z}=\pi$ and 0, respectively, the emergence of the edge states seems to correlate well with the Zak phase. 
This is the same property as in the SSH model. A peculiarity in the network model is that there are two band gaps around $\phi=0$ and $\phi=\pi/2$, and the edge states can emerge in both the gaps. In fact, at  $(\theta_{\rm AB},\theta_{\rm BA})=(0.8\pi,0.6\pi)$ ($\theta_{\rm Z}=\pi$), we have two edge states at the gap around $\phi=\pi/2$. Furthermore, we have four edge states   (two at $\phi=0$ and other two at $\phi=\pi/2$) at 
$(\theta_{\rm AB},\theta_{\rm BA})=(0.4\pi,0.8\pi)$, whereas the Zak phase is zero.

\begin{figure}
\includegraphics[width=0.45\textwidth]{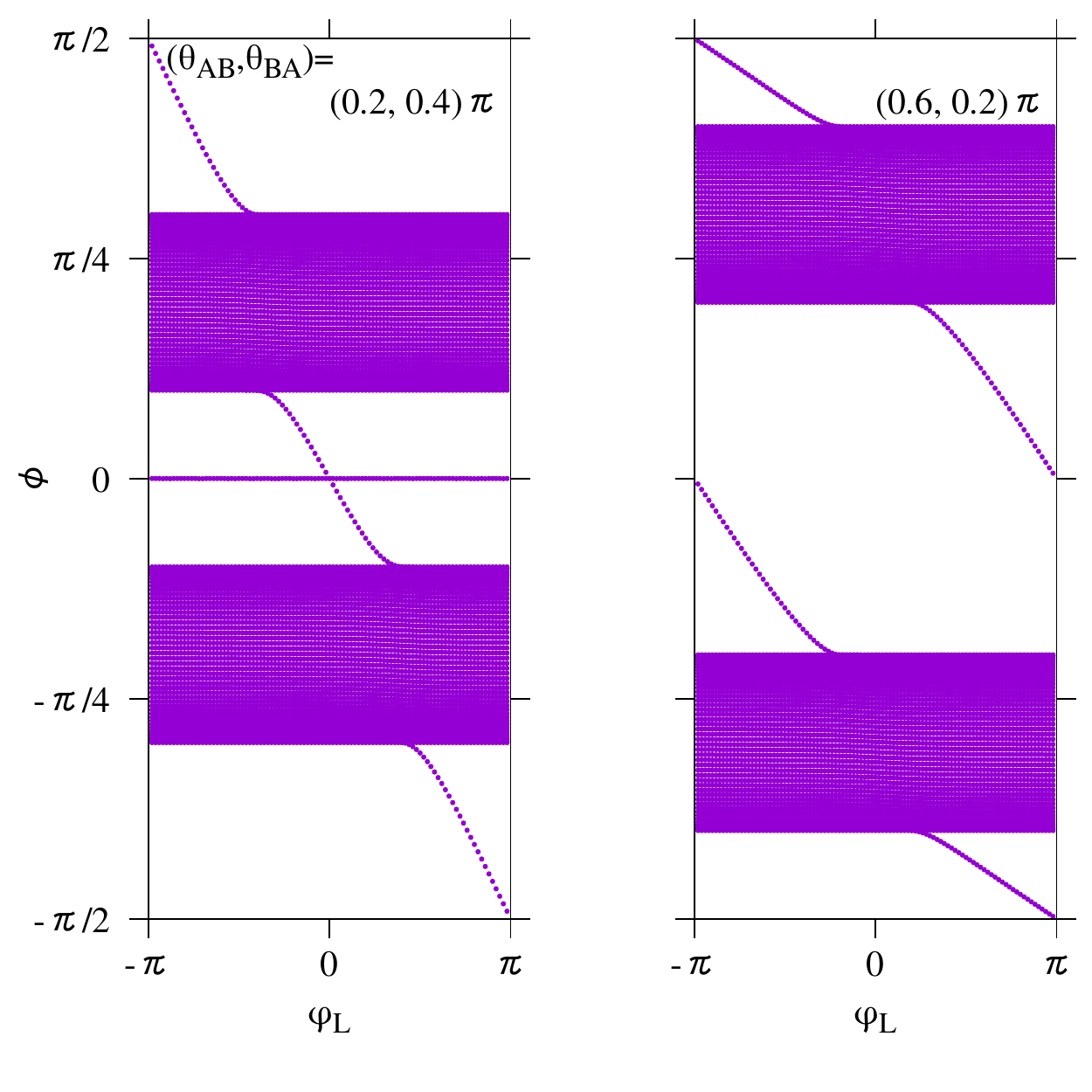}
\caption{\label{Fig_finite1d_psL} Eigenvalues of finite system of $N=50$ at (a) $(\theta_{\rm AB},\theta_{\rm BA})=(0.2\pi,0.4\pi)$  and at (b) $(0.6\pi,0.2\pi)$, as a function of phase delay $\varphi_{\rm L}$ at the left edge. The phase delay at the right edge is kept fixed to $\varphi_{\rm R}=0$. The zero mode that is insensitive to $\varphi_{\rm L}$ in the left panel  is the edge mode localized near the right edge. }
\end{figure}

To further understand edge states in the Floquet network, we consider the response of the edge states against the phase delay. 
We should remind that the results of Fig. \ref{Fig_finite1d} are obtained under a naive boundary condition of zero phase delays, $\varphi_{\rm L}=\varphi_{\rm R}=0$.  
Figure \ref{Fig_finite1d_psL} shows the spectra as a function of the phase  delay at the left edge.  
As we can see, if we include nonzero $\varphi_{\rm L}$, two edge states per fixed $\varphi_{\rm L}$ emerge at the left edge even for the zero Zak phase (the right panel of Fig. \ref{Fig_finite1d_psL}).  One is in the band gap around $\phi=0$, and the other is in the band gap around $\phi=\pi/2$. They disappear around $\varphi_{\rm L}=0$. If the Zak phase is $\pi$ (the left panel of Fig. \ref{Fig_finite1d_psL}), we have only one edge state at the left edge. However, it can emerge also in the band gap around  $\phi=\pi/2$.  The edge state at $\phi=0$ of the right boundary.

A standard understanding in the SSH model is that if the Zak phase is $\pi$, the system exhibits a zero mode at the edge. This is also OK for our model at $\varphi_{\rm L}=0$.  To understand twisted cases ($\varphi_{\rm L}\ne 0$) in our model, we need further criteria.

Let us re-consider the condition of having an edge state in the gaps.  To this end, it is convenient to introduce the S-matrix in the finite system. The S-matrix is defined as 
\begin{align}
\left(\begin{array}{l}
b_N'\\
a_1'
\end{array}\right) = S_N \left(\begin{array}{l}
a_1\\
b_N
\end{array}\right). 
\end{align}
In the gap region, the incident light from the left or right edge cannot propagate in the bulk, and is just reflected. Therefore, the S-matrix behaves like  
\begin{align}
S_N\to \left(\begin{array}{cc}
0 & R'_\infty \\
R_\infty & 0
\end{array}\right) \quad (N\to\infty), \label{Eq_Rinfty}
\end{align} 
with $|R_\infty|=|R'_\infty|=1$ from the unitarity. We thus have 
\begin{align}
&a_1'={\rm e}^{{\rm i}{\rm arg}R_\infty}a_1. 
\end{align}
On the other hand, the boundary condition at the left edge is given by Eq. (\ref{Eq_left_BC}).  
To have an edge-state solution somewhere in the gap, we need 
\begin{align}
{\rm arg}R_\infty(\phi) + 2\phi = -\varphi_{\rm L} \; ({\rm mod}2\pi). \label{Eq_ssh1d_wind}
\end{align}

The semi-infinite reflection matrix $R_\infty$ is available analytically via the forward propagating or forward evanescent eigenmode of the transfer matrix $T(\phi)$ as \cite{Botten:N:M:d:A::6404:p046603:2001}  
\begin{align}
&R_\infty = r_{AB} + t'_{AB}\frac{\beta}{\alpha}, \\
&T(\phi)\left(\begin{array}{cc}
\alpha\\
\beta
\end{array}\right) = \tau\left(\begin{array}{cc}
\alpha\\
\beta
\end{array}\right), \quad |\tau|\le 1 ,\\
&
\left(\begin{array}{cc}
a_{n+1}\\
b_{n+1}
\end{array}\right)= 
T(\phi)\left(\begin{array}{cc}
a_n\\
b_n
\end{array}\right).
\end{align}
In the bulk band regions, $\tau=\exp({\rm i}k_x)$ and the eigenstate is also the eigenstate of $U_{\rm bulk}(k_x)$. In the gap regions, $\log(1/\tau)$ represents the inverse penetration depth of the evanescent mode at propagating phase $\phi$.  
This $R_\infty$ is also related to the Zak phase.
For instance, at the band edge of $k_x=\pi$, $\lambda_\pi=0$ or $\pi$. 
This implies $\beta/\alpha=\pm 1$, so that ${\rm arg}R_\infty=\pm \theta_{\rm AB}$.

\begin{figure}
\centerline{
\includegraphics[width=0.45\textwidth]{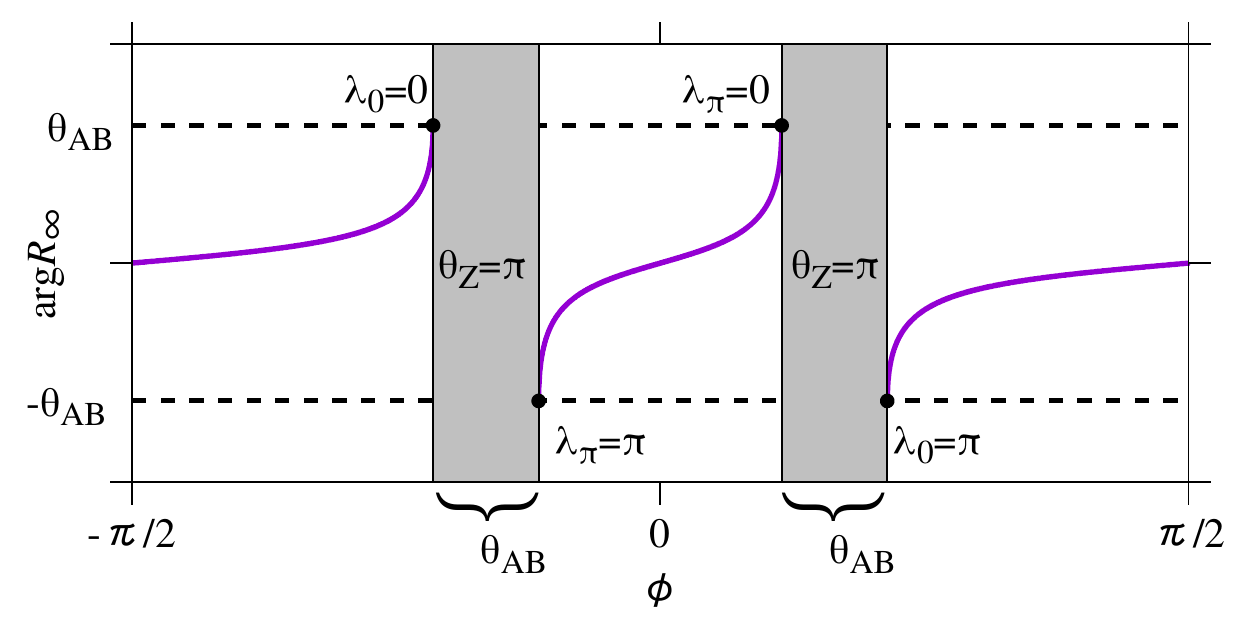}
}
\caption{\label{Fig_bulk1d_argR_schematic} Typical behavior of ${\rm arg}R_\infty$ as a function of $\phi$ in case of $\theta_{\rm Z}=\pi$. Grey region represent the bulk-band regions. Parameter $\lambda$ is related to the semi-infinite reflection coefficient $R_\infty$ and Zak phase $\theta_{\rm Z}$. Here, we also assume that the band width is given by $\theta_{\rm AB}(>0)$. }
\end{figure}

A schematic behavior of ${\rm arg}R_\infty$ as a function of $\phi$ is given in Fig. \ref{Fig_bulk1d_argR_schematic}. 
We found that ${\rm arg}R_\infty$ is an increasing and odd function of $\phi$ in the band gaps. At the band edges, it is pinned to $\pm \theta_{\rm AB}$, depending on the $\lambda$ parameter there. This parameter is related to the Zak phase through Eq. (\ref{Eq_Zak}).

\begin{figure}
\begin{center}
\includegraphics[width=0.45\textwidth]{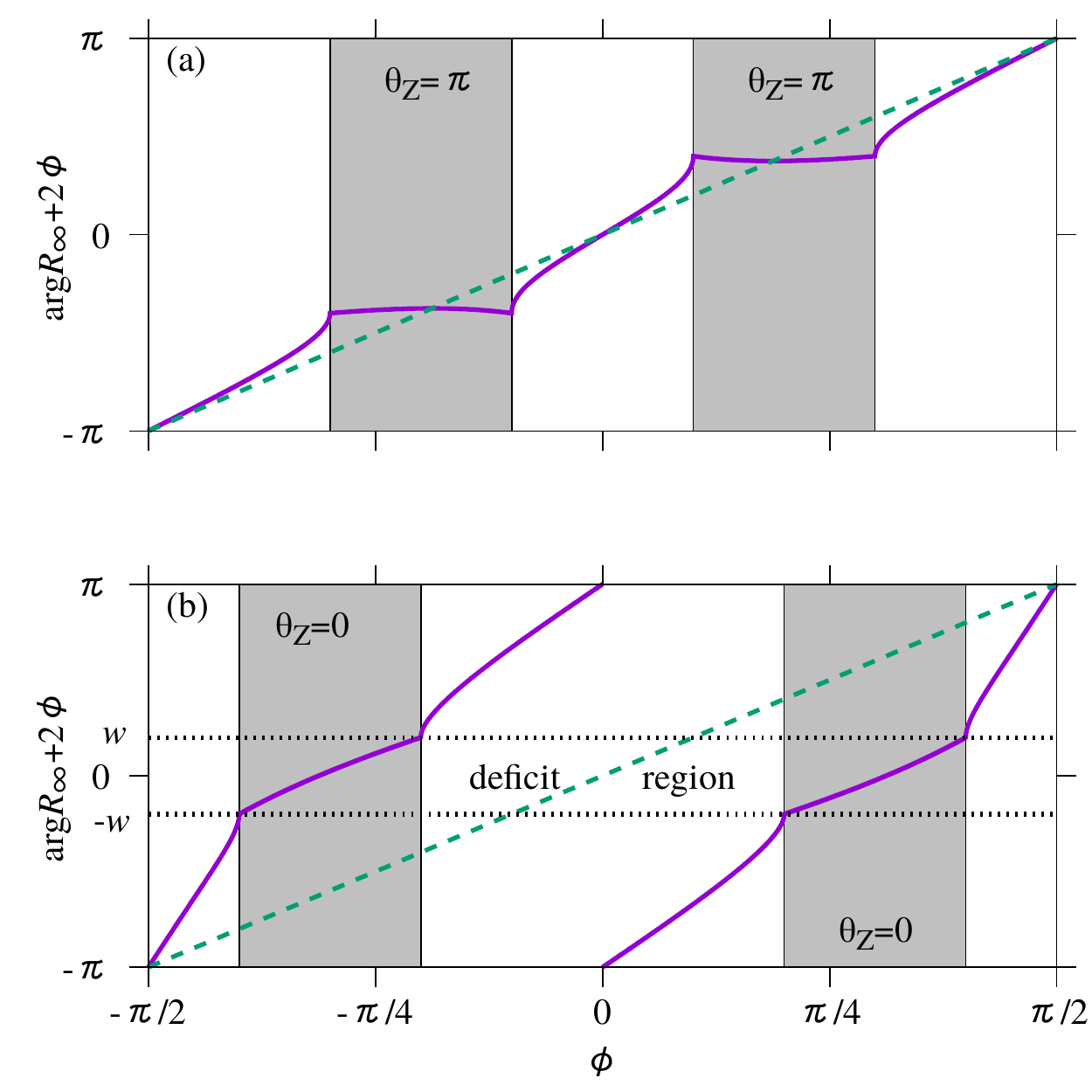}\\
\includegraphics[width=0.45\textwidth]{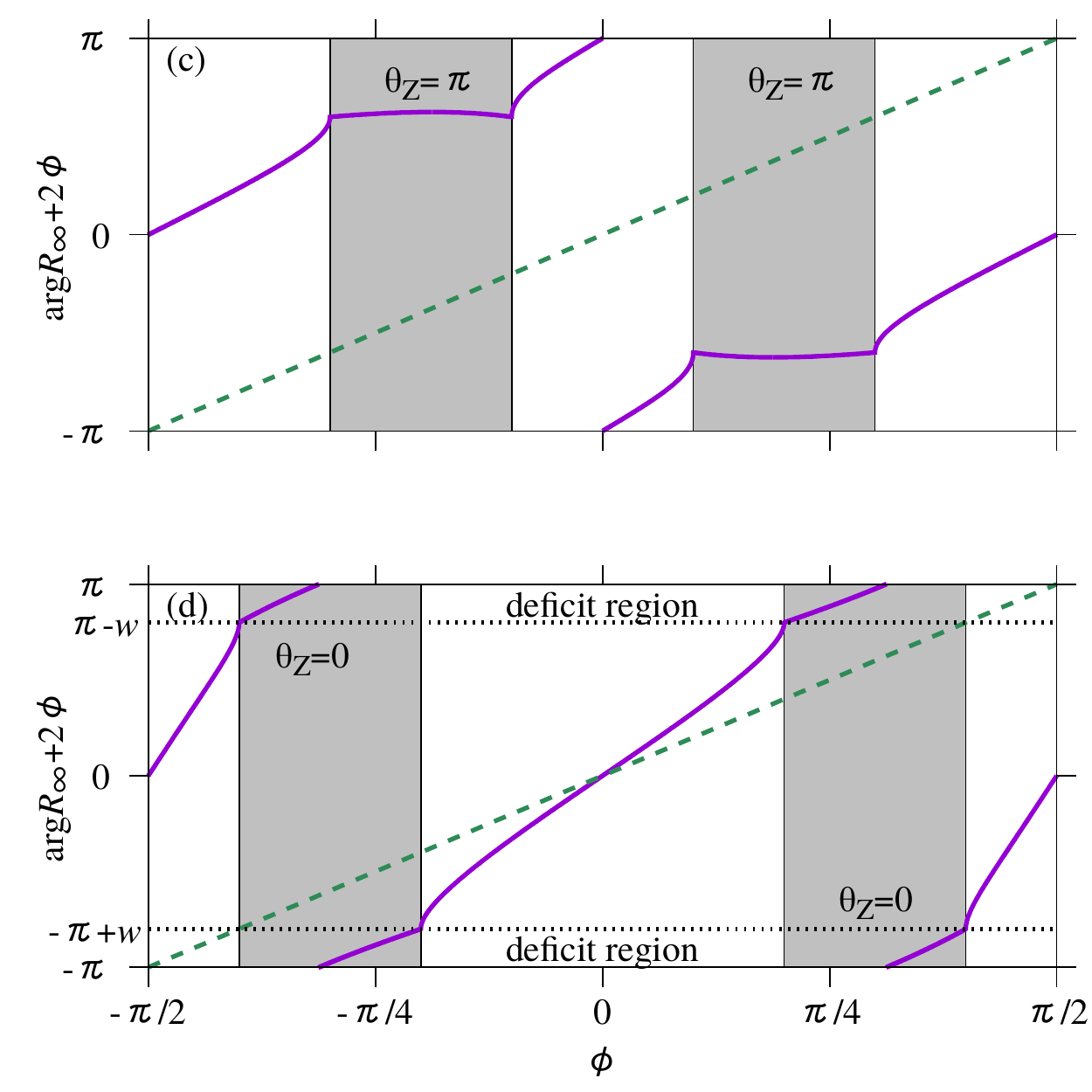}
\end{center}
\caption{\label{Fig_bulk1d_argR} Winding features of ${\rm arg}R_\infty$ as a function of $\phi$. (a) $(\theta_{\rm AB},\theta_{\rm BA})=(0.2,0.4)\pi$, (b)  $(0.6,0.2)\pi$, (c)  $(0.8,0.6)\pi$, (d)  $(0.4,0.8)\pi$.  
Solid line represents ${\rm arg}R_\infty + 2\phi$. Dashed line represents $2\phi$. Grey regions represent the bulk band regions of band width $w$.  Region enclosed by dotted line represents the deficit region such that  
if the phase delay $\varphi_{\rm L}$ is in this region, no edge state emerges.}
\end{figure}

Figure \ref{Fig_bulk1d_argR} shows the left-hand side of Eq. (\ref{Eq_ssh1d_wind}). 
We can see that if the Zak phase is $\pi$, a perfect winding of $2\pi$ as a function of $\phi$ takes place.  Therefore, we have only one edge state irrespective of $\varphi_{\rm L}$. This perfect winding is due to the cancellation of discontinuity (of $2\theta_{\rm AB}$) of ${\rm arg}R_\infty$ in the bulk band regions by the addition of twice the band width $w$ ($2w=2\theta_{\rm AB}$ in (a)) via $2\phi$ term of ${\rm arg}R_\infty(\phi) + 2\phi$. 
On the other hand, if the Zak phase is 0, we have two windings with the deficit of the bulk band regions. The deficit region of $\varphi_{\rm L}$ has $2w$ width centered at $\varphi_{\rm L}=0$ or $\pi$. 
  If the phase delay is in the deficit region, no edge state is found. Otherwise, two edge states emerge. These results explain what happens in Figs. \ref{Fig_finite1d} and \ref{Fig_finite1d_psL}.

\section{Summary}

In summary, we have presented the detailed physical properties of the SSH-type Floquet network.  It can simulate the light propagation in ring-resonator arrays with a dimer-like unit cell. We have characterized the network model by the Zak phase, and found a nontrivial phase diagram. The phase diagram well correlates with the number of the edge states in the band gaps of the quasienergy spectra, regardless of the boundary condition.  A novelty of this system is that even if the Zak phase is 0, the system system can exhibit two edge states per edge. We have presented an analytic criterion for having the two edge states, and confirm it numerically.

\begin{acknowledgments}
This work was supported by JSPS KAKENHI Grant No. 17K05507. 
\end{acknowledgments}


%

\end{document}